\documentclass{amsart}
\usepackage{amssymb}
\usepackage{amsfonts}
\usepackage{amsmath}
\usepackage[legalpaper,bookmarks=true,colorlinks=true,linkcolor=blue,citecolor=blue]{hyperref}
\usepackage{graphicx}%
\usepackage{fancyhdr}
\usepackage{color}
\usepackage[mathlines]{lineno}
\usepackage{lscape}
\usepackage{epsfig}
\usepackage{natbib}
\usepackage{geometry}
\usepackage{tgbonum}
\usepackage[]{listings}

\fontfamily{qcr}\selectfont

\newtheorem{theorem}{Theorem}
\theoremstyle{plain}

\newtheorem{lemma}{Lemma}

\numberwithin{equation}{section}

\newcommand{\Bin}{\bigskip \noindent}

\newcommand{\Ni}{\noindent}

\begin{document}
\Large
\title[]{Moments estimators and omnibus chi-square tests for some usual probability laws}
\author{Gorgui Gning $^{\dag}$}
\author{Aladji Babacar Niang $^{\dag\dag}$}
\author{Modou Ngom $^{\dag\dag\dag}$}
\author{Gane Samb Lo $^{\dag\dag\dag\dag}$}

\begin{abstract} For many probability laws, in parametric models, the estimation of the parameters can be done in the frame of the maximum likelihood method, or in the frame of moment estimation methods, or by using the plug-in method, etc. Usually, for estimating more than one parameter, the same frame is used. We focus on the moment estimation method in this paper. We use the instrumental tool of the functional empirical process (fep) in Lo (2016) to show how it is practical to derive, almost algebraically, the joint distribution Gaussian law and to derive omnibus chi-square asymptotic laws from it. We choose four distributions to illustrate the method (Gamma law, beta law, Uniform law and Fisher law) and completely describe the asymptotic laws of the moment estimators whenever possible. Simulations studies are performed to investigate for each case the smallest sizes for which the obtained statistical tests are recommendable. Generally, the omnibus chi-square test proposed here work fine with sample sizes around fifty.\\

\noindent Gorgui Gning $^{\dag}$\\
LERSTAD, Gaston Berger University, Saint-Louis, S\'en\'egal.\\
Imhotep Mathematical Center\\
Email: gning.gorgui1@ugb.edu.sn, gorguigning003@gmail.com\\

\noindent Aladji Babacar Niang $^{\dag\dag}$\\
LERSTAD, Gaston Berger University, Saint-Louis, S\'en\'egal.\\
Imhotep Mathematical Center\\
Email: niang.aladji-babacar@ugb.edu.sn, aladjibacar93@gmail.com\\

 \noindent Dr Modou Ngom $^{\dag\dag\dag}$ .\\
Work Affiliation : Ministery of High School (SENEGAL)\\
LERSTAD, Gaston Berger University, Saint-Louis, S\'en\'egal
Imhotep Mathematical Center\newline
Email:ngom.modou1@ugb.edu.sn, ngomodoungom@gmail.com\\

\noindent Pr Gane Samb Lo $^{\dag\dag\dag\dag}$ .\\
LERSTAD, Gaston Berger University, Saint-Louis, S\'en\'egal (main affiliation).\newline
LSTA, Pierre and Marie Curie University, Paris VI, France.\newline
AUST - African University of Sciences and Technology, Abuja, Nigeria\\
Imhotep Mathematical Center\\
Email:gane-samb.lo@edu.ugb.sn, gslo@aust.edu.ng, ganesamblo@ganesamblo.net\\
Permanent address : 1178 Evanston Dr NW T3P 0J9,Calgary, Alberta, Canada.\\

\end{abstract}
\maketitle

\section{Introduction}

\Ni Parameter estimations are important steps in parametric statistical modeling. Estimators of parameters can be derived from the maximum likelihood approach, the plug-in methods, the moment methods, etc. Of course, by far Maximum Likelihood Estimators (MLE) are preferred because of the statistical meaning of its derivation.
Moments estimators (\textbf{ME})'s and (\textit{MLE})'s may exists without having closed-form expressions. When a \textit{MLE} does no have a closed-from estimator, the \textit{ME} is a backup solution for authors who wish to a clear idea of  the estimation and a quick and more controlled ways of computation. Finding \textit{ME}'s is an important step in modeling. However,  deriving the related statistical tests is needed for accepting or rejecting hypotheses.\\

\Ni For more that two parameters, it is more practicable to join the individual normal asymptotic laws for each parameters into one chi-square asymptotic laws which is qualified as omnibus following the Jarque-Berra chi-square asymptotic law.\\

\Ni This motivates us to investigate asymptotic laws of \textit{ME}'s estimators of as much as possible of usual and non-usual statistical laws. The found law should be validated by simulation studies before being proposed to potential users.\\

\Ni For a large review of asymptotic estimations and statistical tests, we refer to \cite{vaart_asymp}, \cite{billingsley}, etc. Especially, for methods including functional empirical process, \cite{vaart} is recommended.\\

\Ni However the main tool used here, but not limited to, is the function empirical process (fep) transformed into a instrument tools in \cite{lofep} (see below). let us begin by giving a few words on that tool and next basic notation.\\

\Ni In this paper, we used the \textit{fep} tool to show direct and efficient ways for deriving asymptotic statistical tests for moment estimators for a selected set of four probability laws. Four these laws, all the computations are given in details. Computer codes for simulations are also provided. We could have treated more statistical distributions. However, we wanted this paper to be a model for researchers who need asymptotic statistical tests. Later, we expect to compose a handbook which includes a great number of laws.\\

\Ni The paper is organized as follows. We will close this introductory section by describing the \textit{lofep} tool in Subsection \ref{ss-fep} and, in Section \ref{omnibus} by showing how to derive omnibus chi-square tests from the Gaussian asymptotic theorems for distributions of more than two parameters. In Section \ref{sec2}, we expose the asymptotic laws of the moments estimators of \textit{gamma}, \textit{uniform}, \textit{beta} and \textit{Fisher} distributions. The proofs and the implementation of the \textit{fep} tool on these distributions are stated in Section \ref{sec4}. In Section \ref{sec3}, we proceed to a simulation study on the asymptotic results and show that the omnibus chi-square tests work fine for small samples. The codes used for the simulations are stated in an appendix from page \pageref{scripts_all}. The paper ends with conclusions and perspectives in Section \ref{sec5}.\\

\subsection{A brief reminder of the \textit{fep}} \label{ss-fep} Let $Z_{1}$, $Z_{2}$, ... be a sequence of independent copies of a random variable $Z$ defined on the same probability space with values on some metric space $(S,d)$. Define for each $n\geq 1,$ the functional empirical process by

\begin{equation*}
\mathbb{G}_{n}(f)=\frac{1}{\sqrt{n}}\sum_{i=1}^{n}(f(Z_{i})-\mathbb{E} f(Z_{i})),
\end{equation*}

\Bin where $f$ is a real and measurable function defined on $\mathbb{R}$ such that

\begin{equation}
\mathbb{V}_{Z}(f)=\int \left( f(x)-\mathbb{P}_{Z}(f)\right)
^{2}dP_{Z}(x)<\infty ,  \label{var}
\end{equation}

\Bin which entails

\begin{equation}
\mathbb{P}_{Z}(\left\vert f\right\vert )=\int \left\vert f(x)\right\vert
dP_{Z}(x)<\infty \text{.}  \label{esp}
\end{equation}

\Bin Denote by $\mathcal{F}(S)$ - $\mathcal{F}$ for short -the class of real-valued measurable functions that are defined on S such
that (\ref{var}) holds. The space $\mathcal{F}$ , when endowed with the addition and the external multiplication by real scalars, is a linear space.
Next, it remarkable that $\mathbb{G}_{n}$ is linear on $\mathcal{F}$, that is for $f$ and $g$ in $\mathcal{F}$ and for $(a,b)\in \mathbb{R}{^{2}}$, we have

\begin{equation*}
a\mathbb{G}_{n}(f)+b\mathbb{G}_{n}(g)=\mathbb{G}_{n}(af+bg).
\end{equation*}

\Bin We have this result

\begin{lemma} \label{lemma.tool.1}
\bigskip Given the notation above, then for any finite number of elements $f_{1},...,f_{k}$ of $\mathcal{S},k\geq 1,$ we have

\begin{equation*}
^{t}(\mathbb{G}_{n}(f_{1}),...,\mathbb{G}_{n}(f_{k}))\rightsquigarrow
\mathcal{N}_{k}(0,\Gamma (f_{i},f_{j})_{1\leq i,j\leq k}),
\end{equation*}

\Bin where
\begin{equation*}
\Gamma (f_{i},f_{j})=\int \left( f_{i}-\mathbb{P}_{Z}(f_{i})\right) \left(f_{j}-\mathbb{P}_{Z}(f_{j})\right) d\mathbb{P}_{Z}(x),1\leq i,j\leq k.
\end{equation*}
\end{lemma}

\Bin \textbf{Proof}. It is enough to use the Cram\'{e}r-Wold Criterion (see for example \cite{billingsley}, page 45), that
is to show that for any $a=^{t}(a_{1},...,a_{k})\in \mathbb{R}^{k},$ by denoting $T_{n}=^{t}(\mathbb{G}_{n}(f_{1}),...,\mathbb{G}_{n}(f_{k})),$ we
have $<a,T_{n}>\rightsquigarrow <a,T>$ where $T$ follows the $\mathcal{N}%
_{k}(0,\Gamma (f_{i},f_{j})_{1\leq i,j\leq k})$\ law and $<\circ ,\circ >$
stands for the usual product scalar in $\mathbb{R}^{k}.$ But, by the standard central limit theorem in $\mathbb{R}$, we have

\begin{equation*}
<a,T_{n}>=\mathbb{G}_{n}\left( \sum\limits_{i=1}^{k}a_{i}f_{i}\right)= \sum\limits_{i=1}^{k} a_{i}\mathbb{G}_{n}\left(f_{i}\right)
\rightsquigarrow N(0,\sigma _{\infty }^{2}),
\end{equation*}

\Bin where, for $g=\sum_{1\leq i\leq k}a_{i}f_{i}$,

\begin{equation*}
\sigma _{\infty }^{2}=\int \left( g(x)-\mathbb{P}_{Z}(g)\right) ^{2} \ d\mathbb{P}_{Z}(x)
\end{equation*}

\Bin and this easily gives

\begin{equation*}
\sigma _{\infty }^{2}=\sum\limits_{1\leq i,j\leq k}a_{i}a_{j}\Gamma
(f_{i},f_{j}),
\end{equation*}

\Bin so that $N(0,\sigma _{\infty }^{2})$ is the law of $<a,T>.$ The proof is finished.

\subsection{Main notations in the \textit{fep}} \label{omnibus} In the context of this paper, we use univariate samples $X$, $X_1$, $X_2$, $\cdots$, $X_n$, $n\geq$, with common cumulative distribution function (\textit{cdf}) $F_X=F$, defined on the same probability space $(\Omega, \mathcal{A}, \mathbb{P})$. We will usually need the cumulants $m_k$ and the centered moments $\mu_k$ defined by
$$
m_k=\mathbb{E}X^k, \ \ \mu_k=\mathbb{E} (X-\mathbb{E}(X))^k, \ \ k\geq 1
$$

\Bin and  their plug-in estimators

$$
\overline{X}_{k,n}=\frac{1}{n}\sum_{j=1}^{n} X_{j}^{k}, \ \ \overline{\mu}_{k,n}=\frac{1}{n}\sum_{j=1}^{n} \left(X_{j}-\overline{X}_n\right)^k, \ \ k\geq 1
$$

\Bin with the special case of the empirical mean $\overline{X}_{k,n}=\overline{X}_{n}$. Also the standard variance

$$
S_{n}^2=\frac{1}{n-1}\sum_{j=1}^{n} \left(X_{j}-\overline{X}_n\right)^2
$$

\Bin will be preferred to the plug-in estimator $\overline{\mu}_{2,n}$ of $\sigma^2=\mathbb{V}ar(X)$. We suppose that any moment of order $k\geq 1$ exists whenever it is  used.\\

\Bin The moment method in a parametric estimation related to the studied random variable having $\ell\geq 1$ parameters $(\theta_1,\cdots,\theta_\ell)$ and which generates the sample $\{X_1, \cdots, X_n\}$ consisted in simultaneously solving $\ell$ equations, each of these equations $r \in \{1,\cdots,\ell\}$ being the equality between a cumulant or a moment of order $k_r$ and the corresponding plug-in estimator of the same order $h_r$ where all order $k_j$ are pairwise distinct. In general, it is simpler to take equations between the $\ell^{th}$ first cumulants or moments. The solution, whenever exists and statistics of the empirical cumulant or moments,

$$
\hat \theta_n=(\hat \theta_{1,n},\cdots,\hat \theta_{\ell,n})
$$

\Bin is the vector moment estimator (\textit{ME}).\\

\Ni Once the \textit{ME}'s are found, we will need the joint asymptotic law of the vector $\hat \theta_n$. The tool of the \textit{fep} will greatly help in that target. We will go beyond and derive chi-square tests as much as possible.\\

\Ni The rest of the paper is organized as follows...\\

\subsection{Chi-square law derivation} We are going to show how to derive asymptotic chi-square laws from moment estimators for at least two parameters. In each case below, we treat a two-parameter estimation problem. Suppose that the two parameters are denoted by $a$ and $b$ and their moment estimators are denoted by $\hat a_n$ and $\hat b_n$, $n\geq 2$. We will get in each case a first law in the form: as $n\rightarrow +\infty$,

\begin{equation}
\left(\sqrt{n}(\hat a_n-a), \  \sqrt{n}(\hat b_n-b)\right)^T \rightsquigarrow Z, \ Z\sim\mathcal{N}_2(0,\Sigma), \label{chisq_01}
\end{equation}
 
\Bin where $\sigma_1^2=\Sigma_{1,1}$, $\sigma_2^2=\Sigma_{2,2}$ and $\sigma_{12}=\Sigma_{1,2}$. From usual properties of Gaussian vectors, we have that, whenever 
$det(\Sigma)=\sigma_1^2\sigma_2^2-\sigma_{12}^2\neq 0$,

$$
Z^T \Sigma^{-1} Z \sim \chi_{2}^{2}.
$$

\Bin (See for example \cite{ips-mfpt-ang}, Proposition 12, page 150). By the continuous mapping theorem (see for example \cite{ips-wcia-ang}, Proposition 03, page 34 ), we will have, as $n\rightarrow +\infty$,

$$
Q_n=\left(\sqrt{n}(\hat a_n-a), \  \sqrt{n}(\hat b_n-b)\right) \Sigma^{-1} \left(\sqrt{n}(\hat a_n-a), \  \sqrt{n}(\hat b_n-b)\right)^T \rightsquigarrow  \chi_{2}^{2},
$$

\Bin which, as $n\rightarrow +\infty$, leads to

\begin{equation}
Q_n=\frac{n}{\sigma_1^2\sigma_2^2-\sigma_{12}^2} \left[\sigma_2^2(\hat a_n-a)^2+\sigma_1^2(\hat b_n-b)^2 - 2\sigma_{12} (\hat a_n-a)(\hat b_n-b)\right] \rightsquigarrow \chi_{2}^{2}. \label{chisq_02}
\end{equation}

\Bin So, below, for each treated case, we will state two results according to \eqref{chisq_01} and \eqref{chisq_02}.

\section{Asymptotics related to moments estimators} \label{sec2}

\Ni In that section, we are going to treat the following  probability laws:

$$
(1) X \sim \gamma(a,b), \ \ (2) X \sim \beta (a,b) , \ \ (3) X \sim \mathcal{U} (a,b)  \ \  and \ \  (4) X \sim \mathcal{F} (a,b)
$$

\Bin These results are meant to be interesting examples for other cases not handing here. We stress that the techniques in \cite{lofep} will be extensively used in the following.

\subsection{Gamma laws $\gamma(a,b)$ of parameters $a>0$, $b>0$}

\Ni The gamma law $\gamma(a,b)$ has the probability density function \textit{pdf}

$$
f(x)=\frac{b^a}{\Gamma(a)} x^{a-1} \ \exp(-bx) 1_{(x\geq 0)} \ with \ \Gamma(a)=\int_{a}^{+\infty} x^{a-1} \ \exp(-x) \ dx.
$$
 
\Bin The $k$-th cumulant ($k\geq 1$) is given by

$$
\mathbb{E} X= \frac{a}{b} \  \  \ and \  \  \ \mathbb{E} X^k= \frac{a}{b^k} \prod_{j=1}^{k-1} (a+j) \ for \ k\geq 2,
$$

\Bin and the variance is

$$
\sigma^2=\mathbb{V}ar(X)=\frac{a}{b^2}.
$$

\Bin The moment estimators $\hat a_n$ and $\hat b_n$ are solution of the equations $a/b=\overline{X}_{n}$ and $ab^{-2}=S_n^2$. We get

$$
(\hat a_n, \ \  \hat b_n)=\left(\frac{\overline{X}_{n}^2}{S_n^2}, \ \  \frac{\overline{X}_{n}}{S_n^2}\right).
$$

\Bin Here are the results for the $\gamma$-law of parameters $a>0$ and $b>0$.\\

\begin{theorem}\label{gamma_01} We have

$$
\sqrt{n}(\hat a_n-a, \ \ \hat b_n-b) \rightsquigarrow \mathcal{N}_2(0, \Sigma),
$$

\Bin with

$$
\Sigma_{1,1}=\mathbb{V}ar(H(X)), \ \ \Sigma_{2,2}=\mathbb{V}ar(L(X)), \ \  
\Sigma_{1,2}=\mathbb{C}ov(H(X), L(X))
$$

\Bin and

$$
H=\frac{2\mu \left( \sigma^2 +1\right) }{\sigma ^{4}}h_{1}+\frac{\mu ^{2}}{\sigma ^{4}}h_{2},
$$

$$
L=\left( \frac{\sigma ^{2}+2\mu }{\sigma ^{4}}\right) h_{1}-\frac{\mu }{\sigma ^{4}}h_{2}.
$$

\Bin We also have

\begin{equation}
Q_n=\frac{n}{det(\Sigma)} \left [\Sigma_{2,2}(\hat a_n-a)^2+\Sigma_{1,1}(\hat b_n-b)^2 - 2\Sigma_{1,2} (\hat a_n-a)(\hat b_n-b) \right] \rightsquigarrow \chi_{2}^{2}. \label{chisq_02gamma}
\end{equation}
\end{theorem}

\subsection{ Beta law $\beta(a,b)$ of parameters $a>0$, $b>0$}

\Ni The Beta law has the following probability distribution function

\begin{equation*}
f\left(x\right) =\frac{x^{a-1}\left(1-x\right) ^{b-1}}{B\left( a,b\right) } ,x> 0.
\end{equation*}

\Bin Where

\begin{equation*}
B\left(a,b\right) =\frac{\Gamma \left( a\right) \Gamma \left( b\right) }{\Gamma \left( a+b\right) }.
\end{equation*}

\Bin The expectation is given by

\begin{equation*}
\mathbb{E}\left( X\right) =\frac{a}{a+b}
\end{equation*}

\Bin and the second moment order cumulant is  given by

\begin{equation*}
\mathbb{E}\left(X^2 \right) =\frac{a(a+1)}{\left( a+b\right)\left(a+b+1\right) }.
\end{equation*}

\Bin The moment estimators $\hat a_n$ and $\hat b_n$ are solutions of the equations $a/(a+b)=\overline{X}_{n}$ and $a(a+1)/(a+b)(a+b+1)=\overline{{X}_{n}^2}$. We get

$$
(\hat a_n, \ \  \hat b_n)=\left(\frac{\overline{X}_{n}\left(\overline{X}_{n}-\overline{{X}_{n}^2}\right)}{\overline{{X}_{n}^2}-\overline{X}_{n}^2	}, \ \  \frac{\left(1-\overline{X}_{n}\right)\left(\overline{X}_{n}-\overline{{X}_{n}^2}\right)}{\overline{{X}_{n}^2}-\overline{X}_{n}^2}\right)
$$

\Bin Here are the results for the $\beta$-law of parameters $a>0$ and $b>0$.\\

\begin{theorem}\label{asympt_beta_01} We have

$$
\sqrt{n}(\hat a_n-a, \ \ \hat b_n-b) \rightsquigarrow \mathcal{N}_2(0, \Sigma),
$$

\Bin with

$$
\Sigma_{1,1}=\mathbb{V}ar(H(X)), \ \ \Sigma_{2,2}=\mathbb{V}ar(L(X)), \ \  
\Sigma_{1,2}=\mathbb{C}ov(H(X), L(X)),
$$

$$
H=\frac{\sigma^{2}\left( 2\mu -m_{2}\right) +2\mu^2 \left( \mu-m_{2}\right) }{\sigma ^{4}}h_{1}-\frac{\sigma^{2}\mu  + \mu \left( \mu -m_{2}\right)}{\sigma^{4}}h_{2}
$$

\Bin and

$$
L=\frac{\sigma^{2}\left( m_{2} - 2\mu+ 1 \right)+ 2\mu(1-\mu)(\mu -m_{2} )}{\sigma^{4}}h_{1}+\frac{\left( \mu - 1 \right) \left( \sigma^{2}+ \mu -m_{2}\right)}{\sigma^{4}}h_{2}.
$$

\Bin We also have  the following asymptotic $\chi^2$ result

\begin{equation}
Q_n=\frac{n}{det(\Sigma)} \left[\Sigma_{2,2}(\hat a_n-a)^2+\Sigma_{1,1}(\hat b_n-b)^2 - 2\Sigma_{1,2} (\hat a_n-a)(\hat b_n-b)\right] \rightsquigarrow \chi_{2}^{2}. \label{chisq_02gamma}
\end{equation}

\end{theorem}

\subsection{ The Uniform law of parameters $\mathcal{U}(a,b)$ , $a>0$ and $b>a$}

\Ni The probability distribution function of the uniform law is given by

\begin{equation*}
f\left(x\right) =\frac{1}{b-a}, x\in \left[a,b\right].
\end{equation*}

\Bin Its expectation is
 
\begin{equation*}
\mathbb{E}\left(X\right) =\frac{a+b}{2}.
\end{equation*}

\Bin The variance is

\begin{equation*}
\mathbb{V}ar\left(X\right) =\frac{\left(b-a\right) ^{2}}{12}.
\end{equation*}

\Bin The moment estimators are the solutions of the equations

$$
(\hat a_n, \ \  \hat b_n)=\left(\overline{X}_{n}-\lambda \left( S_{n}^{2}\right) ^{1/2}, \ \  \overline{X}_{n}+\lambda \left(S_{n}^{2}\right) ^{1/2}\right)
$$

\Bin where $\lambda=12^{1/2}/2$.

\Bin Here are the results for the Uniform-law.\\

\begin{theorem}\label{asympt_Uniform_01}  We have

$$
\sqrt{n}(\hat a_n-a, \ \ \hat b_n-b) \rightsquigarrow \mathcal{N}_2(0, \Sigma),
$$

\Bin with

$$
\Sigma_{1,1}=\mathbb{V}ar(H(X)), \ \ \Sigma_{2,2}=\mathbb{V}ar(L(X)), \ \  
\Sigma_{1,2}=\mathbb{C}ov(H(X), L(X)),
$$

\Bin where

$$
H=\left( 1+\frac{\lambda\mu }{\sigma }\right) h_{1}-\frac{\lambda }{2\sigma }h_{2}
$$

\Bin and

$$
L=\left( 1-\frac{\lambda\mu }{\sigma }\right) h_{1}+\frac{\lambda }{2\sigma }h_{2},
$$

\Bin where $\lambda=12^{1/2}/2$.\\

\Bin We also have  

\begin{equation}
Q_n=\frac{n}{det(\Sigma)} \left[\Sigma_{2,2}(\hat a_n-a)^2+\Sigma_{1,1}(\hat b_n-b)^2 - 2\Sigma_{1,2} (\hat a_n-a)(\hat b_n-b) \right] \rightsquigarrow \chi_{2}^{2}. \label{chisq_02gamma}
\end{equation}
\end{theorem}

\subsection{ Fisher law $\mathcal{F}(a,b)$ of parameters $a>0$ and $b>0$}

\Ni For a Fisher law with $a$ and $b$ degrees of freedom, the parameters are supposed to be integers. But in the general case, the probability density function has the same form and is associated to the quotient of two independent random variables $Z_1/Z_2$, where $Z_1 \sim \gamma(a,1/2)/a$ and $Z_2 \sim \gamma(b,1/2)/b$, where $a$ and $b$ are positive. The \textit{pdf} is expressed as follows:

\begin{equation*}
f\left(x\right) =\frac{a^{1/2}b^{1/2}\Gamma \left(\frac{a+b}{2}\right)
x^{a/2-1}}{\Gamma \left(a/2\right) \Gamma \left(b/2\right) \left(b+ax\right) ^{\left(a+b\right) /2}},x>0.
\end{equation*}

\Bin The expectation is given by

\begin{equation*}
\mathbb{E}\left(X\right) =\frac{b}{b-2}=\overline{X_{n}}.
\end{equation*}

\Bin The variance is given by

\begin{equation*}
\mathbb{V}ar\left(X\right) =\frac{2b^{2}\left(a+b-2\right) }{a\left(b-2\right) ^{2}\left(b-4\right) }=S_{n}^{2}, \ b>4.
\end{equation*}

\Bin The moment estimators are solutions of the equations

\begin{equation*}
\left( \hat{a}_{n},\ \ \hat b_{n}\right) =\left(\frac{2\overline{X}_{n}^{2}}{S^2(2-\overline{X}_{n})-\overline{X}_{n}^{2}(\overline{X}_{n}-1)}, \ \ \frac{2\overline{X}_{n}}{\overline{X}_{n}-1}\right).
\end{equation*}

\Bin Here are the results for the Fisher-law.\\

\begin{theorem}\label{asympt_Fisher_01} We have

$$
\sqrt{n}(\hat a_n-a, \ \ \hat b_n-b) \rightsquigarrow \mathcal{N}_2(0, \Sigma),
$$

\Bin with

$$
\Sigma_{1,1}=\mathbb{V}ar(H(X)), \ \ \Sigma_{2,2}=\mathbb{V}ar(L(X)), \ \  
\Sigma_{1,2}=\mathbb{C}ov(H(X), L(X)),
$$

\Bin where
$$
H=\frac{2\mu(2-\mu)}{\beta}h_{1}-\frac{2\mu^{2}(2-\mu)}{\beta^{2}}h_{2}
$$

\Bin with

$$
\beta=\sigma^{2}+2\mu(2-\mu)-\mu(3\mu-2)
$$

\Bin and where 

$$
L=-\frac{2}{(\mu -1)^2}h_{1}.
$$

\Bin We also have  

\begin{equation}
Q_n=\frac{n}{det(\Sigma)} \left[ \Sigma_{2,2}(\hat a_n-a)^2+\Sigma_{1,1}(\hat b_n-b)^2 - 2\Sigma_{1,2} (\hat a_n-a)(\hat b_n-b)\right]\rightsquigarrow \chi_{2}^{2}. \label{chisq_02gamma}
\end{equation}
\end{theorem}

\section{Simulations} \label{sec3}

\Bin We are going to describe our simulation works for one of studied distributions. Next we will explain their outputs and their interpretations. Finally, we will display results for all cases. Important scripts will be posted in the appendix \pageref{script_01}.\\

\subsection{Simulation works}. In all cases, we estimate two parameters. In the case of the $\gamma(a,b)$ law, the moment estimators are denoted by \textit{achap} and and \textit{bchap}. We will have three parts.\\

\Ni \textbf{A- Computing the exact moments and other coefficients of the estimators}.\\

\Ni (a) Before proceeding to the Monte-Carlo method, we have to computed the function $H$ and $L$, demoted as \textit{bigH} and \textit{bigL}.\\

\Ni (b) We proceed to numerical methods for computing $\mathbb{E}H(X)$, $sigmaHexaC=\mathbb{V}ar(E)H(X)$, $\mathbb{E}L(X)$, 
$sigmaLexaC=\mathbb{V}ar(E)L(X)$ and the exact co-variance $SigmaHLexa=\mathbb{C}ov(H(X),L(X)$, where $X$ stands for random variable with the studied law (here a $\gamma(a,b)$ law). The trapezoidal method algorithm is used for all integral computations here.\\

\Ni In page \pageref{script_01}, the related script is given under the title \textbf{A1 - Computing exact coefficients} . \Bin Table \label{tab1} gives exact values of the variances for different pairs $(a,b)$.\\

\begin{table}
	\centering
	\label{tab1}
	\caption{Over/under estimation of variances and covariances of estimators with respect to the true values for $(a,b)=(2,3))$}
		\begin{tabular}{l|lll}
		 \hline \hline
		(a,b) & (2,3) & (3,10)& (10,3)\\
			\hline \hline
	\textit{sigmaHexaC}   	&  	$7.78$    	&  $107.08$    	&  $63.08$     \\
	\textit{sigmaLexaC}		&  	$7.37$    	&  $76.58$     	&  $16.99$     \\
	\textit{sigmaHLexa}  		&  	$40.3$    	&  $7.985.01$  	&  $1006.95$     \\
	\textit{Correlation}		&		$69.76\%$  	& $98.11\%$ 		&$93.95\%$\\
	\hline \hline
\end{tabular}
\end{table}

\Ni \textbf{B- Monte-Carlo estimation}.\\

\Ni (a) Fix a sample size $n\geq 2$. Fix values to $a$ and $b$.\\

\Ni (b) Fix the number of repetitions $B=1000$ (big enough to ensure the stability of outcomes).\\

\Ni (c) At each repetition $j \in \{1,\cdots,B\}$, we generate an sample of $X$ of size $n$. Next
\begin{itemize}
\item[(1)] $DA[j]=\sqrt{n}(achap-a)$
\item[(2)] $DB[j]=\sqrt{n}(bchap-a)$
\item[(3)] $VH[j]=sd(bigH(X))$
\item[(4)] $VL[j]=sd(bigL(X)$
\item[(5)] $VHL[j]=cov(bigH(X),bigL(X)$
\end{itemize} 

\Ni In page \pageref{script_02}, the related script is under the title \textbf{A2- Script Monte Carlo works}.\\

\Ni \textbf{C- Computing the empirical moments and other coefficients}.\\

\Ni (a) Now, we have: (1) an estimate of $\Sigma_{1,1}$ by the square of the average of the vector $VH$, (2) an estimate of $\Sigma_{2,2}$ by the square of the average of the vector $VH$ and (3) an estimate of $\Sigma_{1,2}$ by  the average of the vector $VHL$. We denote them as \textit{SigmaHEMP}, \textit{SigmaLEMP} and \textit{SigmaHLEMP}.\\

\Ni (b) We also have: (1) an estimate of $\Sigma_{1,1}$ by empirical variance of $DA$, (2) an estimate of $\Sigma_{2,2}$ 
by empirical variance of $DB$ and  (3) an estimate of $\Sigma_{1,2}$ by  by empirical covariance between $DA$ and $DB$. We denote them as \textit{sigmaHSAMP}, \textit{sigmaLSAMP} and \textit{sigmaHLSAMP}.\\

\Ni In page \pageref{script_03}, the related script for computing \textit{sigmaHEMP} , \textit{sigmaLEMP}, \textit{sigmaHLEMP}, \textit{sigmaHSAMP}, \textit{sigmaLSAMP} and \textit{sigmaHLSAMP} is given under the title \textbf{A3- Over/under estimations of variances and covariances} from the script \textbf{A2- Script Monte Carlo works} in page  \pageref{script_02}.\\

\Ni  In Tables \ref{tab2a}, \ref{tab2b} and \ref{tab2c} display the quotients of empirical coefficients over the true coefficients, allowing to over or under-estimation, for three values of pairs $(a,b)$.\\  

\begin{table}
\centering
\caption{Over/under estimation of variances and covariances of estimators with respect to the true values for $(a,b)=(2,3))$}
\label{tab2a}
\begin{tabular}{l|llll}
\hline \hline
size 					& n=50 					& n=100 				& n=200 				& n=1000\\
\hline \hline
\textit{Qsig-1emp} 		& $98.98\%$    	&  $98.12\%$   	&$98.47\%$   		& $98.49\%$    \\
\textit{Qsig-2emp} 		& $72.00\%$     &  $74.59\%$    &$74.85\%$   		&$75.56\%$    \\
\textit{Qsig-12emp} 		& $81\%$     		&  $81.05\%$    &$82.39\%$			&   $81.89\%$    \\
\textit{Qsig-1samp} 		& $44.03\%$     &  $43.34\%$    &$43.57\%$   		& $44.93\%$    \\
\textit{Qsig-2samp} 		& $79.11\%$     &  $76.62\%$    &$76.58\%$   		&$78.81\%$    \\
\textit{Qsig-12samp} 	& $46.34\%$     &  $44.57\%$    &$45.04\%$   		&$48.23\%$    \\
\hline \hline
\end{tabular}
\end{table}

\begin{table}
\centering
\caption{Over/under estimation of variances and covariances of estimators with respect to the true values for $(a,b)=(3,10))$}
\label{tab2b}
\begin{tabular}{l|lllll}
\hline \hline
size & n=50 & n=100 & n=200 & n=1000 \\
\textit{Qsig-1emp} & $96.87\%$    &  $97.59\%$     &   $98.46\%$  &   $98.19\%$  \\
\textit{Qsig-2emp} & $99.44\%$     &  $99.95\%$     &   $100.49\%$  &   $100.32\%$  \\
\textit{Qsig-12emp} & $98.72\%$     &  $99.33\%$     &   $100.46\%$  &   $100.00\%$   \\
\textit{Qsig-1samp} & $4.96\%$     &  $4.46\%$     &   $4.43\%$  &   $4.40\%$    \\
\textit{Qsig-2samp} & $25.03\%$     &  $22.34\%$     &   $21.98\%$  &   $22.12\%$    \\
\textit{Qsig-12samp} & $1.199\%$     &  $0.96\%$     &   $0.93\%$ &   $0.94\%$   \\
\hline \hline
\end{tabular}
\end{table}

\begin{table}
\centering
\caption{Over/under estimation of variances and covariances of estimators with respect to the true values for $(a,b)=(10,3))$}
    \label{tab2c}
    \begin{tabular}{l|lllll}
    \hline \hline
size & n=50 & n=100 & n=200 &n=1000\\
\textit{Qsig-1emp} & $91.58\%$    &  $92.92\%$     &   $94.29\%$  &  $94.25\%$  \\
\textit{Qsig-2emp} & $90.25\%$     &  $91.77\%$     &   $93.30\%$   &  $94.29\%$  \\
\textit{Qsig-12emp} & $85.72\%$     &  $86.95\%$     &   $88.93\%$  &  $88.14\%$  \\
\textit{Qsig-1samp} & $25.09\%$     &  $23.26\%$     &   $23.79\%$  &  $23.54\%$  \\
\textit{Qsig-2samp} & $30.30\%$     &  $28.15\%$     &   $28.19\%$   &  $28.37\%$ \\
\textit{Qsig-12samp} & $7.43\%$     &  $6.404\%$     &   $6.56\%$   &  $6.54\%$  \\
\hline \hline
\end{tabular}
\end{table}

\Ni \textbf{D- Statistical tests for Computing the empirical moments and other coefficients}.\\

\Ni (1) Performance of the point estimation. From the script \textbf{A2- Script Monte Carlo works} in page  \pageref{script_02}, we can compute the mean error (ME), the mean absolute error (MAE) and the square-root of the mean square error (MSE)of the point estimations  on $a$ and $b$ the R codes \textit{mean(DACHAP-a)},
\textit{mean(DBCHAP-b)}, \textit{mean(abs(DACHAP-a))}. \textit{mean(abs(DBCHAP-b))}, \textit{sd(DACHAP-a)} and \textit{sd(DBCHAP-b)}. We report their values in Table \ref{tabpe}

\begin{table}
	\centering
		\begin{tabular}{cccccccc}
			Error type & n=25 & n=50 & n=75 & n=100 & n=200 & n=300 & n=1000\\
			\hline \hline
			ME (A) & 1.09		& 	0.53& 	0.33	&0.13 	& 0.13		&  0.08  &0.02\\
			MAE (A)  & 2.88		& 1.99	& 1.54		& 0.92	& 0.92		&  0.76  &0.411\\
$\sqrt{MSE}$ (A)  & 3.88		&2.49 	& 1.99		& 1.163	& 	1.17	&  0.95  &0.5\\
			\hline \hline
			ME (B)  & 0.42		& 0.21	& 0.12		& 0.05	& 0.048		&  0.029  &0.009\\
			MAE (B)  & 1.07		& 0.72	& 0.57		& 0.34	& 0.034		&  0.28  &0.15\\
$\sqrt{MSE}$ (B)&1.44 		& 	0.92&0.73 		& 0.43	& 	0.43	&  0.35  &0.19\\
			\hline \hline
		\end{tabular}
	\caption{Evolution of the errors on estimation of $a=10$ and $b=3$ in the sample sizes}
	\label{tabpe}
\end{table}

\Ni (2) Statistical tests on $a$. We have three tools

\begin{eqnarray*}
&&DA_1[j]=\sqrt{\frac{n}{SigmaHexaC}} (achap-a)\approx \mathcal{N}(0,1),\\
&&DA_2[j]=\sqrt{\frac{n}{SigmaHEMP}} (achap-a) \approx \mathcal{N}(0,1),\\
&&DA_3[j]=\sqrt{\frac{n}{SigmaHSAMP}} (achap-a) \approx \mathcal{N}(0,1).
\end{eqnarray*}

\Ni (3) Statistical tests on $b$. We have three tools

\begin{eqnarray*}
&&DB_1[j]=\sqrt{\frac{n}{SigmaLexaC}} (bchap-b) \approx \mathcal{N}(0,1),\\
&&DB_2[j]=\sqrt{\frac{n}{SigmaLEMP}} (bchap-b)\approx \mathcal{N}(0,1),\\
&&DB_3[j]=\sqrt{\frac{n}{SigmaLSAMP}} (bchap-b) \approx \mathcal{N}(0,1).
\end{eqnarray*}

\Ni In page \pageref{script_04}, the scripts \textbf{A4- Computations of the p-values}, for each parameter $a$ and $b$, we compute the empirical p-values for each sequence, as the frequency of element of the sequence exceeding $1.96$. The test is satisfactory if that $p$ value is less of around $5\%$. The different p-values for $a=2$ and $b=3$ are given for different values of $n$ in Table \ref{tab3c}.

\begin{table}
\centering
\caption{Empirical p-values for statistical test of $a$ and $b$ ($a=10$ and $b=3$)}
\label{tab3c}
\begin{tabular}{l|lllll}
\hline \hline
cases  & n=50 & n=100 & n=150 &n=200 &n=1000\\
Exact    &  $0\%$    &  $0\%$     &  $0\%$  &  $0\%$  & $0\%$  \\
Empirical &  $0\%$    &  $0\%$     &  $0\%$    &  $0\%$ &  $0\%$ \\
Sample   &  $5.4\%$    &  $5.3\%$     &  $5.18\%$  &  $5.09\%$ &  $5\%$   \\
Exact & $0.01\%$    &  $0\%$     &   $0\%$  &  $0\%$  &  $0\%$ \\
Empirical & $0.02\%$     &  $0\%$     &   $0\%$   &  $0\%$ &  $0\%$ \\
Sample & $5.43\%$     &  $5.49\%$     &   $5.12\%$  &  $5.28\%$ &  $5.28\%$ \\
\hline \hline
\end{tabular}
\end{table}

\begin{figure}[h]
	\centering
		\includegraphics[width=0.60\textwidth]{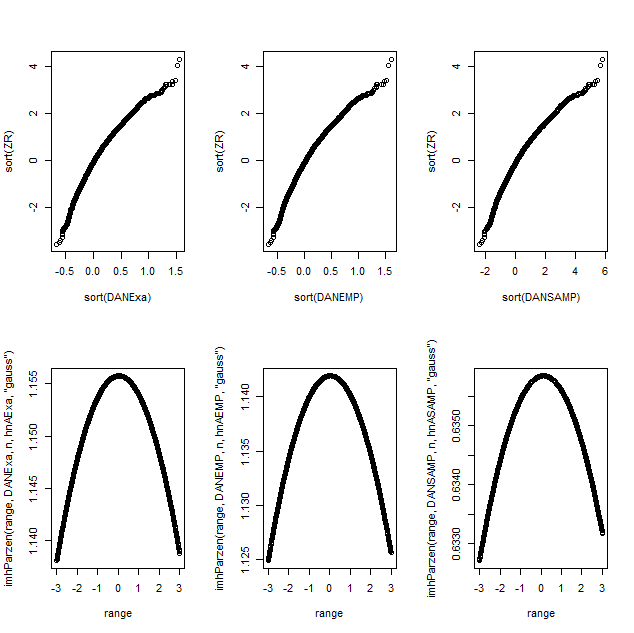}
	\caption{QQ-plots and Parzen estimators related to the parameter a for n=50, according to the type of estimation of the coefficients}
	\label{fig3a50}
\end{figure}

\begin{figure}[h]
	\centering
		\includegraphics[width=0.60\textwidth]{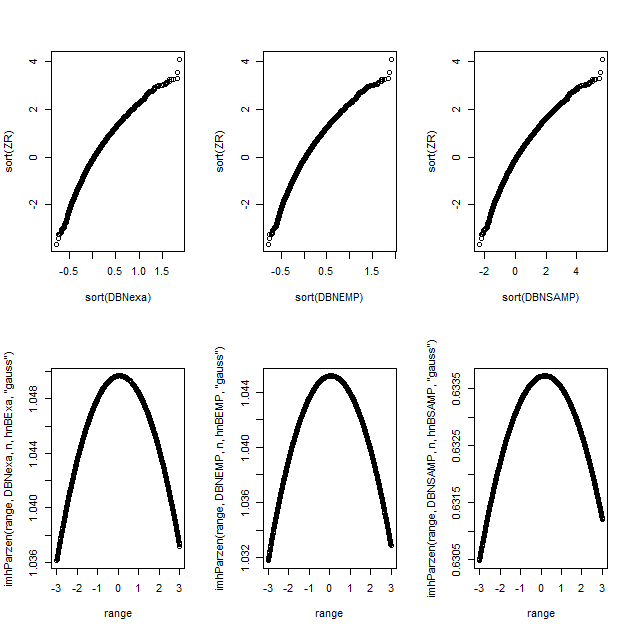}
	\caption{QQ-plots and Parzen estimators related to the parameter a for n=50, according to the type of estimation of the coefficients}
	\label{fig3b50}
\end{figure}

\begin{figure}[h]
	\centering
		\includegraphics[width=0.60\textwidth]{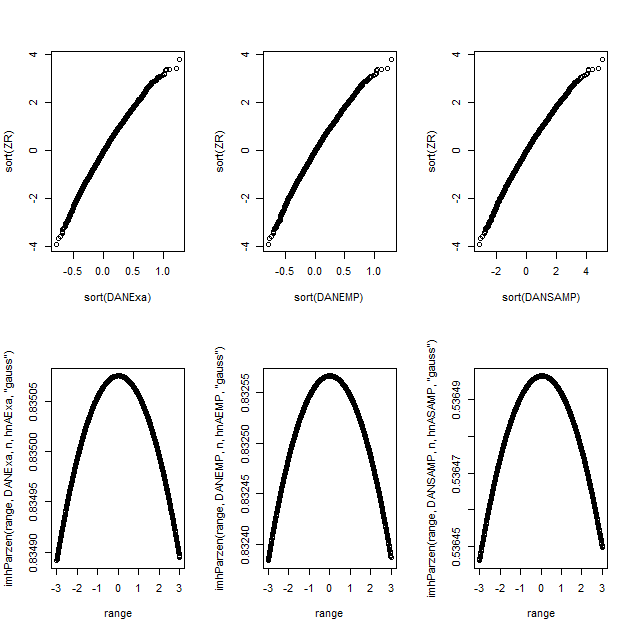}
	\caption{QQ-plots and Parzen estimators related to the parameter a for n=300, according to the type of estimation of the coefficients}
	\label{fig3a300}
\end{figure}

\begin{figure}[h]
	\centering
		\includegraphics[width=0.60\textwidth]{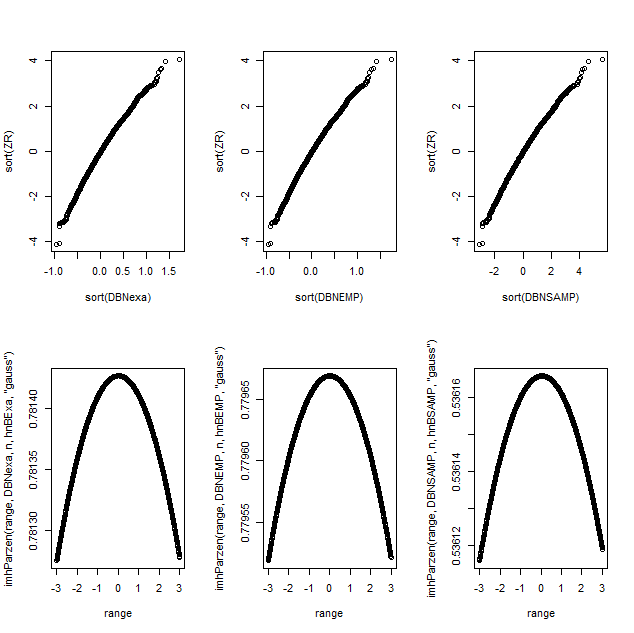}
	\caption{QQ-plots and Parzen estimators related to the parameter a for n=300, according to the type of estimation of the coefficients}
	\label{fig3b300}
\end{figure}

\Ni To test the quality of the normal approximations, we display the QQ-plots and the Parzen estimators graphs for each parameter in Fig 
\ref{fig3a50} (QQ-plots and Parzen estimators related to the parameter $a$ for n=50, according to the type of estimation of the coefficients),
in Fig \ref{fig3b50} (QQ-plots and Parzen estimators related to the parameter $b$ for n=50, according to the type of estimation of the coefficients),in Fig \ref{fig3a300} (QQ-plots and Parzen estimators related to the parameter $a$ for n=300, according to the type of estimation of the coefficients) and in Fig \ref{fig3b300} (QQ-plots and Parzen estimators related to the parameter $b$ for n=300, according to the type of estimation of the coefficients) in Appendix C
(Page \pageref{appendixC}).\\

\Bin \textbf{(E) Omnibus test}. We mean by omnibus test that the combine both test into a chi-square test as in Part (b) of each of Theorems \ref{asympt_Fisher_01}, \ref{asympt_Uniform_01}, \ref{asympt_beta_01} \label{gamma_01}. Depending on the use of exact values, empirical values or sample values of the variance of co-variances, we have
three statistics that can be used each for the chi-square test:

\begin{eqnarray*}
\text{QExa}[j]&=&\frac{n}{detHL} \biggr(sigmaLExaC(achap[j]-a)^2\\
			&+&sigmaHexaC(bchap[j]-b)^2\\
			&-& 2\times sigmaHLexa(achap[j]-a)\times (bchap[j]-b)\biggr) \sim \chi_1^2\\
\text{QEMPDB}[j]&=&\frac{n}{detHLEMP} \biggr((sigmaLEMP^2)(achap[j]-a)^2\\
				&+&(sigmaHEMP^2)(bchap[j]-b)^2\\
				&-& 2\times sigmaHLEMP(achap[j]-a)\times (bchap[j]-b)\biggr)  \sim \chi_1^2\\
\text{QEMPDB}[j]&=&\frac{n}{detHLSAMP} \biggr((sigmaLSAMP^2)(achap[j]-a)^2\\
				&+&(sigmaHSAMP^2)(bchap[j]-b)^2\\
				&-&2 \times sigmaHLSAMP(achap[j]-a)\times (bchap[j]-b)\biggr)  \sim \chi_1^2
\end{eqnarray*}

\Bin Table \ref{tab4} provides the p-value related to the omnibus test for different sizes according to the estimations of the coefficients used. The related test is given in the script under the title \textit{A5- p-values for the omnibus statistical test} in page \pageref{script_05}.

\begin{table} 
\centering
\caption{QNE, QNEMP and QNSAMP according to the size $n$ for $a=10$ and $b=3$}
\label{tab4}
\begin{tabular}{l|llllll}
\hline \hline
cases  & 	n=50    & n=100    		&n=200     	&n=300   		&n=1000\\
QNE    &  $1.2\%$   	&  $0.1\%$  	&  $0.02\%$ &  $0.001\%$ &  $0.01\%$ \\
QNEMP  &  $0\%$    &  $0\%$     	&  $0\%$    &  $0\%$    &   $0.8\%$ \\
QNSAMP &  $0.6\%$  	&  $0.1\%$   	&  $0.3\%$  &  $0.01\%$  &  $0.2\%$   \\
\hline \hline
\end{tabular}
\end{table}

\Bin \textbf{(D) - Conclusions of recommendations from simulations}. The simulation studies show that the omnibus statistical test is very good even for sizes as small as $n=50$ for all estimations of the coefficients in the test statistics. When we do Gaussian separate tests for $a$ and $b$, the outcomes are remarkable in the use the variance and covariance of $\sqrt{n}(\hat a_n-a)$ and $\sqrt{n}(\hat b_n-b)$. This is observable in the QQ-plots, the Parzen graphs and in the p-values of the tests. The separate tests seem to recommend the tests when $n$ is bigger than $100$. But, definitively, the omnibus works fine for small sizes as $n=11$ with p-values $5.6\%$, $0\%$, $1.7\%$.\\

\Ni We strongly suggest to no use the tests with empirical estimations of the variance and covariance which lead to severe under or over estimation.

\section{Proofs of Theorems} \label{sec4} 

\Ni Here, we provide the computations for each treated probability law.

\subsection{ Gamma Law $ \gamma(a,b) $ of parameters $a>0$, $b>0$ }
\Ni We have  

\begin{equation}\label{gamma_01}
\overline{X}_{n} =\mu +n^{-1/2}G_{n}\left( h_{1}\right)
\end{equation}

\Bin and

\begin{eqnarray}\label{gamma_02}
S_{n}^{2} &=&\frac{1}{n-1}\left[\sum_{j=1}^{n}X_{j}^{2}-n\overline{X}_n^{2}\right] \\
&=&\frac{n}{n-1}\left[ \frac{1}{n}\sum_{j=1}^{n}X_{j}^{2}-\overline{X}_{n}^{2}\right]\\
&=& \frac{n}{n-1}\left[\overline{X_{n}^{2}} - \overline{X}_{n}^{2}\right].
\end{eqnarray}

\Bin By the delta method,

\begin{equation}\label{gamma_03}
\overline{X}_n^{2} =\mu ^{2}+n^{-1/2}G_{n}\left( 2\mu h_{1}\right) +O_{p}\left(n^{-1/2}\right)
\end{equation}
 \Bin and 
\begin{equation}\label{gamma_04}
\overline{X_{n}^{2}}=\frac{1}{n}\sum_{j=1}^{n}X_{j}^{2} =m_{2}+n^{-1/2}G_{n}\left(h_{2}\right).
\end{equation}

\Bin We know that

\begin{equation*}
\frac{n}{n-1}=\left( \frac{n-1}{n}\right) ^{-1}=\left( 1-1/n\right)^{-1}=1+O_{p}\left( 1\right).
\end{equation*}

\Bin Hence

\begin{equation}\label{gamma_05}
S_{n}^{2}=\sigma ^{2}+n^{-1/2}G_{n}\left( h_{2}-2\mu h_{1}\right)+O_{p}\left( n^{-1/2}\right).
\end{equation}

\Bin Let us handle $\hat{b}_{n}$. We have

\begin{equation*}
\hat b_{n}=\frac{\overline{X_{n}}}{S_{n}^{2}}.
\end{equation*}

\Bin By  equations \eqref{gamma_01}, \eqref{gamma_05} and by lemma $11$ in \cite{lofep}, we have

\begin{eqnarray*}
\hat{b_{n}} &=&\frac{\mu }{\sigma ^{2}}+n^{-1/2}G_{n}\left( \frac{h_{1}}{\sigma
^{2}}-\frac{\mu }{\sigma ^{4}}\left( h_{2}-2\mu h_{1}\right) \right)+O_{p}\left( n^{-1/2}\right) \\
&=&\frac{\mu }{\sigma ^{2}}+n^{-1/2}G_{n}\left( L\right),
\end{eqnarray*}

\Bin where

\begin{equation*}
L=\frac{h_{1}}{\sigma ^{2}}-\frac{\mu }{\sigma ^{4}}\left( h_{2}-2\mu h_{1}\right) =\left( \frac{\sigma ^{2}+2\mu h_{1}}{\sigma ^{4}}\right) h_{1}-\frac{\mu }{\sigma ^{4}}h_{2}
\end{equation*}

\Bin Let us treat $\hat a_{n}$. By combining  equations $\eqref{gamma_03}$, $\eqref{gamma_05}$ and lemma $11$ in  \cite{lofep} , we get

\begin{equation*}
\hat a_{n}=\frac{\mu }{\sigma ^{2}}+n^{-1/2}G_{n}\left(H\right),
\end{equation*}

\Bin where

\begin{eqnarray*}
H &=&\frac{2\mu }{\sigma ^{2}}h_{1}-\frac{\mu ^{2}}{\sigma ^{4}}\left(h_{2}-2\mu h_{1}\right) \\
&=&\frac{2\mu \sigma ^{2}+2\mu }{\sigma ^{4}}h_{1}-\frac{\mu ^{2}}{\sigma
^{4}}h_{2} \\
&=&\frac{2\mu \left( \sigma^2 +1\right) }{\sigma ^{4}}h_{1}-\frac{\mu ^{2}}{\sigma ^{4}}h_{2}.
\end{eqnarray*}

\subsection{Beta Law $ \beta (a,b) $ of parameters $a>0$, $b>0 $}

\Ni The moment estimators $\hat a_n$ and $\hat b_n$ are solutions of the equations

\begin{equation*}
\hat{a}_{n}=\frac{\overline{X}_{n}\left( \overline{X}_{n}-\overline{X_{n}^{2}}\right)}{\overline{X_{n}}^{2}-\overline{X_{n}^{2}}}
\end{equation*}

\Bin and

\begin{equation*}
\hat b_{n}=\frac{\left( 1-\overline{X}_{n}\right) \left( \overline{X}_{n}-\overline{X_{n}^{2}}\right) }{\overline{X_{n}}^{2}-\overline{X_{n}^{2}}}.
\end{equation*}

\Bin We have

\begin{equation}  \label{beta_01}
\overline{X}_{n} =\mu +n^{-1/2}G_{n}\left( h_{1}\right)
\end{equation}

\Bin and

\begin{equation}  \label{beta_02}
\overline{X_{n}^{2}} =m_{2}+n^{-1/2}G_{n}\left( h_{2}\right).
\end{equation}

\Bin By the delta method, we have

\begin{equation}  \label{beta_03}
\overline{X_{n}}^2 =\mu ^{2}+n^{-1/2}G_{n}\left( 2\mu h_{1}\right)
+O_{p}\left(n^{-1/2}\right)
\end{equation}

\Bin and

\begin{eqnarray}  \label{beta_04}
1-\overline{X}_{n}&=&1-\mu -n^{-1/2}G_{n}\left( h_{1}\right)  \notag \\
&=&1-\mu +n^{-1/2}G_{n}\left( -h_{1}\right).
\end{eqnarray}

\Bin By combining equations \eqref{beta_01} and \eqref{beta_02}, we have

\begin{equation*}
A_{n}=\overline{X}_{n}-\overline{X_{n}^{2}}=\mu -m_{2}+n^{-1/2}G_{n}\left(h_{1}-h_{2}\right).
\end{equation*}

\Bin By combining equations \eqref{beta_02} and \eqref{beta_03}, we have

\begin{eqnarray*}
B_{n} &=&\overline{X_{n}^2}-\overline{X_{n}}^2=m_{2}-\mu
^{2}+n^{-1/2}G_{n}\left(h_{2}\right) -n^{1/2}G_{n}\left( 2\mu h_{1}\right) \\
&=&m_{2}-\mu ^{2}+n^{-1/2}G_{n}\left( h_{2}-2\mu h_{1}\right)
+O_{p}\left(n^{-1/2}\right).
\end{eqnarray*}

\Bin Hence

\begin{equation*}
\hat a_{n}=\frac{C_{n}}{B_{n}}
\end{equation*}

\Bin with

\begin{eqnarray*}
C_{n} &=&\overline{X}_{n}\times A_{n} \\
&=&\mu \left( \mu -m_{2}\right) +n^{-1/2}G_{n}\left( \left( \mu h_{1}-\mu h_{2}\right) +\left( \mu -m_{2}\right) h_{1}\right) +O_{p}\left(
n^{-1/2}\right).
\end{eqnarray*}

\Bin Then

\begin{equation*}
C_{n}=\mu \left( \mu -m_{2}\right) +n^{-1/2}G_{n}\left(H_{2}\right)+O_{p}\left( n^{-1/2}\right),
\end{equation*}

\Bin where
\begin{eqnarray*}
H_{2} &=&\left( \mu h_{1}-\mu h_{2}\right) +\left( \mu -m_{2}\right) h_{1} \\
&=&\left( 2\mu -m_{2}\right) h_{1}-\mu h_{2}.
\end{eqnarray*}

\Bin Hence

\begin{equation*}
\hat a_{n}=\frac{\mu \left( \mu -m_{2}\right) +n^{-1/2}G_{n}\left(H_{2}\right)+O_{p}\left( n^{-1/2}\right) }{m_{2}-\mu
^{2}+n^{-1/2}G_{n}\left( h_{2}-2\mu h_{1}\right) +O_{p}\left(n^{-1/2}\right) }.
\end{equation*}

\Bin By lemma $11$ in \cite{lofep}, we get

\begin{equation*}
\hat a_{n}=\frac{\mu \left( \mu -m_{2}\right) }{m_{2}-\mu ^{2}}+n^{-1/2}G_{n}\left( H\right) +O_{p}\left( n^{-1/2}\right),
\end{equation*}

\Bin Where

$$
H=\frac{\sigma^{2}\left( 2\mu -m_{2}\right) +2\mu^2 \left( \mu-m_{2}\right) }{\sigma ^{4}}h_{1}-\frac{\sigma^{2}\mu  + \mu \left( \mu -m_{2}\right)}{\sigma^{4}}h_{2}.
$$

\Bin Let us handle now $\hat b_{n}$ .\newline

\Ni Remind that

\begin{equation*}
\hat{b}_{n}=\frac{\left( 1-\overline{X}_{n}\right) \left( \overline{X}_{n}-X_{n}^{2}\right) }{\overline{X_{n}^{2}}-\overline{X}_{n}^{2}}=\frac{B_{1}\left(
n\right) }{B_{2}\left( n\right) }.
\end{equation*}

\Bin  Thanks to the previous calculus, we have

\begin{equation*}
\hat{b}_{n}=\frac{( 1-\mu )(\mu - m_{2})}{\sigma^{2}} +n^{-1/2}G_{n}\left( L\right) +O_{p}\left(n^{-1/2}\right),
\end{equation*}

\Bin where

$$
L=\frac{\sigma^{2}\left( m_{2} - 2\mu+ 1 \right)+ 2\mu(1-\mu)(\mu -m_{2} )}{\sigma^{4}}h_{1}+\frac{\left( \mu - 1 \right) \left( \sigma^{2}+ \mu -m_{2}-\right)}{\sigma^{4}}h_{2}.
$$

\subsection{Uniform Law $\mathcal{U}(a,b)$, of parameters $a>0$ and $b>a$}

\Ni We have

\begin{equation} \label{beta_01}
\overline{X}_{n}=\mu +n^{-1/2}G_{n}\left(h_{1}\right)
\end{equation}

\Bin and

\begin{equation}  \label{beta_02}
\overline{X_{n}^{2}} =m_{2}+n^{-1/2}G_{n}\left( h_{2}\right).
\end{equation}

\Bin By combining  equations \eqref{beta_01} and \eqref{beta_02}, we have

\begin{equation*}
S_{n}^{2}=\sigma ^{2}+n^{-1/2}G_{n}\left( h_{2}-2\mu h_{1}\right).
\end{equation*}

\Bin By the delta method, we have
 
\begin{equation*}
\left( S_{n}^{2}\right) ^{1/2}=\sigma +n^{-1/2}G_{n}\left( \frac{1}{2\sigma}\left( h_{2}-2\mu h_{1}\right) \right) +O_{p}\left(n^{-1/2}\right).
\end{equation*}

\Bin So

\begin{eqnarray*}
a_{n} &=&\mu -\lambda \sigma +n^{-1/2}G_{n}\left( \left( h_{1}-\frac{\lambda}{2\sigma}\left( h_{2}-2\mu h_{1}\right) \right) \right) +O_{p}\left(n^{-1/2}\right) \\
&=&\mu -\lambda \sigma +n^{-1/2}G_{n}\left( L\right) +O_{p}\left(n^{-1/2}\right),
\end{eqnarray*}

\Bin with

\begin{equation*}
L=\left( 1+\frac{\lambda\mu }{\sigma}\right) h_{1}-\frac{\lambda }{2\sigma}h_{2}.
\end{equation*}

\Bin By using the same technique, we have

\begin{equation*}
\hat b_{n}=\mu -\lambda \sigma+n^{-1/2}G_{n}\left( H\right) +O_{p}\left(n^{-1/2}\right),
\end{equation*}

\Bin where

\begin{equation*}
H=\left( 1-\frac{\lambda\mu }{\sigma}\right) h_{1}+\frac{\lambda }{2\sigma}h_{2}.
\end{equation*}

\subsection{Fisher Law $\mathcal{F}(a,b)$ of parameters $a$ and $b$}

\Bin The moment estimators are defined below. The first moment estimator is 

\begin{eqnarray}\label{fisher_01}
\hat{b}&=&\frac{2\overline{X}_{n}}{\overline{X}_{n}-1}\\
       \notag &=&\frac{2\mu+n^{-1/2}\mathbb{G}_{n}(2h_{1})}{\mu-1+n^{-1/2}\mathbb{G}_{n}(-h_{1})}\\
			\notag &=&\frac{2\mu}{\mu-1}+n^{-1/2}\mathbb{G}_{n}(L)+o_{\mathbb{P}}(n^{-1/2}),
\end{eqnarray}

\Bin where

\begin{eqnarray*}
L&=&\frac{2h_{1}}{\mu -1}-\frac{2\mu}{(\mu-1)^2}h_{1}\\
 &=&-\frac{2}{(\mu -1)^2}h_{1}.
\end{eqnarray*}

\Bin The second estimator is given by 

$$
a=\frac{2b^{2}(b-2)}{S^2(b-2)^2(b-4)-2b^4}.
$$

\Bin By equation \eqref{fisher_01}, we know that 

$$
b=\frac{2\overline{X}_{n}}{\overline{X}_{n}-1}.
$$

\Bin So we have,

\begin{eqnarray*}
b-2&=&\frac{2\overline{X}_{n}}{\overline{X}_{n}-1}-2=\frac{2}{\overline{X}_{n}-1},\\
b-4&=&\frac{2\overline{X}_{n}}{\overline{X}_{n}-1}-4=\frac{-2\overline{X}_{n}+4}{\overline{X}_{n}-1}.
\end{eqnarray*}

\Bin Hence 

\begin{eqnarray*}
\hat{a}&=&\frac{16\overline{X}_{n}^{2}}{8S^2(2-\overline{X}_{n})-8\overline{X}_{n}^{2}(\overline{X}_{n}-1)}\\
       &=&\frac{2\overline{X}_{n}^{2}}{S^2(2-\overline{X}_{n})-\overline{X}_{n}^{2}(\overline{X}_{n}-1)}.
\end{eqnarray*}

\Bin From \eqref{fisher_01}, we have

\begin{equation}\label{fisher_02}
S_{n}^{2}=\sigma^{2}+n^{-1/2}\mathbb{G}_{n}(h_{2}-2\mu h_{1})+o_{\mathbb{P}(n^{-1/2})}.
\end{equation} 

\Bin Further, we have 
 
\begin{equation}\label{fisher_03}
2-\overline{X}_{n}=2-\mu +n^{-1/2}\mathbb{G}_{n}(-h_{1})
\end{equation}

\begin{equation}\label{fisher_04}
S_{n}^{2}(2-\overline{X}_{n})=\sigma^{2}(2-\mu)+n^{-1/2}\mathbb{G}_{n}(L_{1})+o_{\mathbb{P}(n^{-1/2})},
\end{equation}

\Bin where 

\begin{eqnarray*}
L_{1}&=&-\sigma^{2}h_{1}+(2-\mu)(h_{2}-2\mu h_{1})\\
  &=&(\sigma^{2}+2\mu(2-\mu))h_{1}+(2-\mu)h_{2}
\end{eqnarray*}

\Bin and

$$
\overline{X}_{n}^{2}=\mu^{2}+n^{-1/2}\mathbb{G}_{n}(2\mu h_{1})+o_{\mathbb{P}(n^{-1/2})}.
$$

\Bin We also have

\begin{equation}\label{fisher_05}
\overline{X}_{n}-1=\mu -1 +\mathbb{G}_{n}(h_{1})
\end{equation}

\begin{equation}\label{fisher_06}
\overline{X}_{n}^{2}(\overline{X}_{n}-1)=\mu ^{2}(\mu -1)+ n^{-1/2}\mathbb{G}_{n}(L_{2}),
\end{equation}

\Bin where 

\begin{equation*}
L_{2}=\mu^{2}h_{1}+(\mu -1)(2\mu h_{1})=\mu(3\mu-2)h_{1}.
\end{equation*}		

\Bin So the denominator $\eqref{fisher_04}$/$\eqref{fisher_06}$ is expanded as

$$
denom=(2-\mu)\sigma^{2}-\mu^{2}(\mu-1)+n^{-1/2}\mathbb{G}_{n}(L_{3})+o_{\mathbb{P}(n^{-1/2})},
$$

\Bin where 

$$
L_{3}=[\sigma^{2}+2\mu(2-\mu)-\mu(3\mu-2)]h_{1}+(2-\mu)h_{2}.
$$

\Bin Hence

\begin{eqnarray*}
\hat{a}&=&\frac{2\mu^{2}+n^{-1/2}\mathbb{G}_{n}(4\mu h_{1})+o_{\mathbb{a}(n^{-1/2})}}{Denom}\\
       &=&\frac{2\mu^{2}}{(2-\mu)\sigma^{2}-\mu^{2}(\mu -1)+n^{-1/2}\mathbb{G}_{n}(H)+o_{\mathbb{a}(n^{-1/2})}},
\end{eqnarray*}

\Bin where 

$$
H=\frac{2\mu}{(2-\mu)\sigma^{2}-\mu^{2}(\mu -1)}h_{1}-\frac{2\mu^{2}}{((2-\mu)\sigma^{2}-\mu^{2}(\mu -1))^{2}}L_{3}.
$$

\Bin Let us set 

\begin{eqnarray*}
\alpha&=&(2-\mu)\sigma^{2}-\mu^{2}(\mu -1)\\
\beta&=&\sigma^{2}+2\mu(2-\mu)-\mu(3\mu-2).
\end{eqnarray*}

\Bin Then 

$$
L_{3}=\beta h_{1}+(2-\mu)h_{2}.
$$

\Bin Hence

\begin{eqnarray*}
\notag H&=&(\frac{4}{\beta}\mu-\frac{2\mu^{2}}{\beta})h_{1}-\frac{2\mu^{2}(2-\mu)}{\beta^{2}}h_{2}\\
 &=&\frac{2\mu(2-\mu)}{\beta}h_{1}-\frac{2\mu^{2}(2-\mu)}{\beta^{2}}h_{2}. \ \ \ \blacksquare
\end{eqnarray*}

\Bin 
\section{Conclusions and perspectives} \label{sec5} Moment estimators for four statistical distributions been studied through their asymptotic Gaussian laws with the help of the \textit{fep} tool. Chi-square omnibus tests have been derived for each distribution. The results have been simulated and the chi-square tests revealed themselves efficient for small sample sizes. The R codes of the simulations are attached to the paper in an appendix. The main perspective is to develop a full chapter in  with the study of a large number of distributions.

\textbf{Acknowledgment}. The authors Niang and Ngom express their thanks to Professor Lo for guidance, and moral and financial assistance.\\

\newpage
\Ni \textbf{Appendix: Scripts}. \label{scripts_all}\\

\Ni \textbf{A1 - Computing exact coefficients} \label{script_01}

\normalsize
\begin{lstlisting}
a=2; b=3
(m1=a/(a+b))
(m2=a*(a+1)/((a+b)*(a+b+1)))
mu=a/b
sigmac=a/(b^2)
(a1=2*mu*(sigmac+1)/(sigmac^2))
(a2=-mu^2/(sigmac^2))
b1=(sigmac+2*mu)/(sigmac^2)
b2=-mu/(sigmac^2)

bigH <- function(x) a1*x+a2*x^2
bigL <- function(x) b1*x+b2*x^2

# calcul variance exacte avec l'algorithme Integrations
#appel de la fonction gamma
CC=(b^a)/gamma(a)

#Variance exacte
#bigHFV <- function(x) x^(a-1)*exp(-b*x)*bigH(x)*CC
#bigLFV <- function(x) x^(a-1)*exp(-b*x)*bigL(x)*CC
#bigHFVC <- function(x) x^(a-1)*exp(-b*x)*(bigH(x)^2)*CC
#bigLFVC <- function(x) x^(a-1)*exp(-b*x)*(bigL(x)^2)*CC
 
bigHInv <- function(u) bigH(qgamma(u,a,b))
bigHInvC <- function(u) bigH(qgamma(u,a,b))^2
bigLInv <- function(u) bigL(qgamma(u,a,b))
bigLInvC <- function(u) bigL(qgamma(u,a,b))^2
bigHLInv<- function(u) bigH(qgamma(u,a,b))*bigL(qgamma(u,a,b))

(moH1=imhIntegD1(0,1,bigHInv,100,0.0001))
(m2H1=imhIntegD1(0,1,bigHInvC,100,0.0001))
(sigmaHexaC=m2H1 - (moH1^2))

(moL1=imhIntegD1(0,1,bigLInv,100,0.0001))
(m2L1=imhIntegD1(0,1,bigLInvC,100,0.0001))
(sigmaLexaC=m2L1 - (moL1^2))
(moHL=imhIntegD1(0,1,bigHLInv,100,0.0001))
(sigmaHLexa=moHL-(moH1*moL1))
(detHL=(sigmaHexaC*sigmaLexaC)-(sigmaHLexa^2))
\end{lstlisting}

\newpage
\Ni \textbf{A2- Script Monte Carlo works}. \label{script_02}\\

\begin{lstlisting}
n=50
B=1000
DACHAP<-numeric(B)
DBCHAP<-numeric(B)
sigmaVL<-numeric(B)
sigmaVH<-numeric(B)
sigmaVHL<-numeric(B)

for(j in 1:B){	
	X=rgamma(n,a,b)
	m1X=mean(X)
	varX=var(X)
	(achap=((m1X)^2)/varX)
	(bchap=((m1X))/varX)
	sigmaVL[j]=sd(bigL(X))
	sigmaVH[j]=sd(bigH(X))
	sigmaVHL[j]=cov(bigH(X),bigL(X))
	DACHAP[j]=achap
	DBCHAP[j]=bchap
}

DA=sqrt(n)*(DACHAP-a)
DB=sqrt(n)*(DBCHAP-b)
\end{lstlisting}

\newpage
\Ni \textbf{A3- Over/under estimations of variances and covariances} \label{script_03}\\

\Bin

\begin{lstlisting}
sigmaHEMP/sqrt(sigmaHexaC)
sigmaLEMP/sqrt(sigmaLexaC)
sigmaHLEMP/sigmaHLexa

sigmaHSAMP/sqrt(sigmaHexaC)
sigmaLSAMP/sqrt(sigmaLexaC)
sigmaHLSAMP/sigmaHexa
\end{lstlisting}

\Bin \textbf{A4- Computations of the p-values for statistical tests pour $a$ et $b$} \label{script_04}.

\Bin

\begin{lstlisting}
imhPvalue(DA/sqrt(sigmaHexaC),B,1.96)
imhPvalue(DA/sigmaHEMP,B,1.96)
imhPvalue(DA/sigmaHSAMP,B,1.96)

imhPvalue(DB/sqrt(sigmaLexaC),B,1.96)
imhPvalue(DB/sigmaLEMP,B,1.96)
imhPvalue(DB/sigmaLSAMP,B,1.96)

See Function imhPvalue, under the title:
B1-(a) Empirical tests from approximated Gaussian data, below 
\end{lstlisting}

\newpage
\Bin \textbf{A5- p-values for the omnibus statistical test} \label{script_05}.\\

\Bin 
\begin{lstlisting}
#detHL calcule dans la partie
#\textbf{A1 - Computing exact coefficients}

#Estimation du determinannt Empirique et Sample
(detHLEMP=(sigmaHEMP^2)*(sigmaLEMP^2)-(sigma12EMP^2))
(detHLSAMP=((sigmaHSAMP^2)*(sigmaLSAMP^2))-(sigma12SAMP^2))

#Calcul du khi-deux avec les parametres exacts
QNE=(sigmaLexaC*(achap-a)^2)+(sigmaHexaC*(bchap-b)^2)
QNE=QNE-2*sigmaHLexa*(achap-a)*(bchap-b)
QNE=QNE/(detHL)

#Calcul du khi-deux avec les parametres empiriques
QNEMP=(sigmaLEMP^2*(achap-a)^2)+(sigmaHEMP^2*(bchap-b)^2)
QNEMP=QNEMP-2*sigma12EMP*(achap-a)*(bchap-b)
QNEMP=QNEMP/(detHLEMP)

#Calcul du khi-deux avec les parametres estimes sur les statistiques
QNSAMP=(sigmaLSAMP^2*(achap-a)^2)+(sigmaHSAMP^2*(bchap-b)^2)
QNSAMP=QNSAMP-2*sigma12SAMP*(achap-a)*(bchap-b)
QNSAMP=QNSAMP/(detHLSAMP)

#mean p-values using function in:
#B2-(b) Empirical tests from approximated Chi-square data

pchisq(QNE,2)
pchisq(QNEMP,2)
pchisq(QNSAMP,2)
\end{lstlisting}

\newpage
\Bin \textbf{B- Auxiliary Functions}.\\

\Ni \textbf{B1-(a) Empirical tests from approximated Gaussian data}. \label{script_B_01}\\

\begin{lstlisting}
size=1000
Z=rnorm(size, 0,1)
pt=1.96
j=0
pv=0

imhPvalue <- function(Z,size,pt){
	pv=0
	for(j in 1:size){
		if( abs(Z[j]) > pt){	
			pv=pv+1
		}
	}
	pv=100*pv/size
	return(pv)
}
\end{lstlisting}

\Ni \textbf{B2-(b) Empirical tests from approximated Chi-square data}. \label{script_B_02}\\

\begin{lstlisting}
size=1000
df=2 // degrees of freedom
Z=rchisq(size, df)
th=0.05
j=0
pv=0

imhPvalueChisq <- function(Z,size,df,th){
	pv=0
	for(j in 1:size){
		if( Z[j] > qchisq(1-th,df)){	
			pv=pv+1
		}
	}
	pv=100*pv/size
	return(pv)
\end{lstlisting}

\Ni \textbf{Appendix C} \label{appendixC}

\end{document}